\documentclass[twocolumn,twoside,slac_two]{revtex4}
\usepackage{graphicx}
\usepackage{fancyhdr}
\usepackage{hyperref}
\hypersetup{
    colorlinks=true,
    urlcolor=magenta,
}
\begin{document}

\title{Precision measurement of the (e$^+$ + e$^-$) flux in primary cosmic rays from 0.5 GeV to 1 TeV with the Alpha Magnetic Spectrometer on the International Space Station}

%

\author{M. Vecchi on behalf of the AMS-02 Collaboration}
\affiliation{Instituto de F\'isica de S\~ao Carlos (IFSC), Universidade de S\~ao Paulo, CP 369, 13560-970,  S{\~a}o Carlos, SP,~Brazil}

\begin{abstract}
We present a precise measurement of the combined electron plus positron flux from 0.5
GeV to 1 TeV, based on the analysis of the data collected by the Alpha Magnetic
Spectrometer  during the first 30 months of operations aboard the International Space Station. The statistics and the high resolution
of AMS-02 detector provide a precise measurement of the flux. The flux is smooth and
reveals new and distinct information. Above
30.2 GeV, the combined electron plus positron flux can be described accurately by a single power law.
\end{abstract}

\maketitle

\thispagestyle{fancy}

\section{Introduction and detector layout}
\label{intro}
\begin{figure*}[ht]
\centering
\includegraphics[width=90mm]{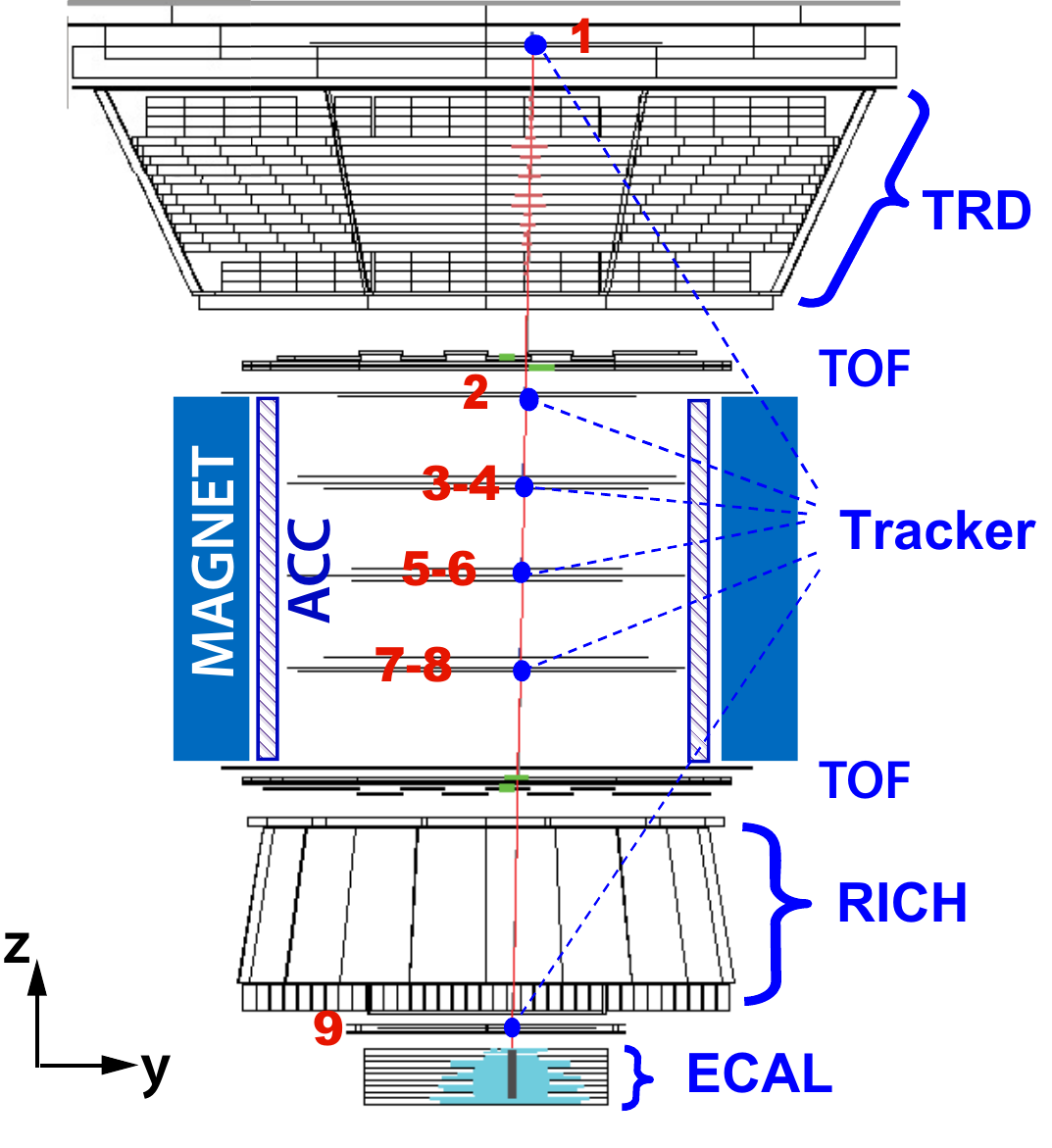}
  \caption{ A 369 GeV positron event as measured by the AMS detector on the ISS, in the (y-z) plane.}
  \label{vecchi:fig1}
\end{figure*}

The Alpha Magnetic Spectrometer is a general purpose particle physics detector, operating in space since May 2011. It will achieve a unique 
long duration mission, aiming at performing antimatter and dark matter searches, as well as cosmic ray composition and flux measurements~\cite{bibSdellato}. 
The experiment is installed onboard the International 
Space Station (ISS), that follows a Low Earth Orbit at about 400 km 
altitude with respect to the Earth surface, well located to detect cosmic particles before they interact with the outer layers of the 
atmosphere. This makes the ISS one the most interesting environments to perform cosmic rays studies. The measurements presented in 
this document are based on the data collected during the first 30 months of operations of the detector, from May 19th 2011 to November 26th 2013. 
In this period  41$\times 10^9$ cosmic ray events were detected. 
The detector is composed of several sub-detectors, as shown in figure~\ref{vecchi:fig1}.
The silicon tracker~\cite{bibTRK} measures the trajectory and absolute charge $|Z|$ of cosmic rays by performing multiple measurements of the coordinates and energy loss.
Together with the 0.14~T permanent magnet, the tracker measures the particle rigidity $R=pc/Ze$, where $p$ is the momentum.
The Transition Radiation Detector (TRD)~\cite{bibTRD} identifies the particle as an electron/positron.
The four layers of the Time of Flight (TOF)~\cite{bibTOF} measure the particles charge and ensure that the particle is downward-going.
The high efficiency ($\sim 99.999\%$) anti-coincidence counters~\cite{bibACC} inside the magnet bore are used to reject particles outside the geometric acceptance.
The Ring Imaging CHerenkov detector (RICH)~\cite{bibRICH} measures the particles charge and velocity.
The imaging Electromagnetic Calorimeter 
(ECAL)~\cite{bibECAL} identifies the particle as an electron/positron and measures its energy.\\
The AMS-02 detector has been extensively calibrated 
using a test beam at CERN with $e^{-}$ and $e^{+}$ from 10 to 290 GeV, with protons at 180 and 400 GeV,
and with $\pi^{\pm}$ from 10 to 180 GeV.
\subsection{Lepton-hadron separation}\label{prejec}
Electrons and positrons only account for a tiny fraction of the cosmic rays: $e^{-}$ are $\sim 10^{-2}$ less abundant than protons, while $e^{+}$ are $\sim 10^{ -4}$ less abundant than protons. However, the measurement of their fluxes can provide important informations to understand the nearby universe, as their detection horizon is limited to few kiloparsecs, due to energy losses. 
Three main sub-detectors provide clean and redundant identification of positrons and electrons with independent suppression of the proton background. 
These are the TRD (above the magnet), the ECAL (below the magnet) and the tracker. 
The matching of the ECAL energy and the momentum measured with the tracker (\textit{E/p} in the following) greatly improves the proton rejection. 
To differentiate between $e^{\pm}$ and protons in the TRD, an estimator formed by the ratio of the log-likelihood probability of the $e^{\pm}$ hypothesis to that of the proton hypothesis in each layer is used. 
The proton rejection power of the TRD estimator at 90$\%$ $e^{\pm}$ efficiency is 10$^{3}$ to 10$^{4}$~\cite{PF2013}, as estimated using flight data.
To cleanly identify electrons and positrons in the ECAL, a Boosted Decision Tree estimator
is built using the 3D shower shape. The ECAL proton rejection 
power reaches 10$^4$ when combined with the \textit{E/p} matching requirement. 
\section{The combined ($e^+$ + $e^-$) flux measurement}\label{allefluxes}
The  data collected during the first 30 months of operations of the detector were analysed to provide 
precise measurements of the positron fraction~\cite{PF2014} and the individual positron and electron fluxes~\cite{epflux2014}.
The positron flux have been measured up to 500 GeV
and of the electron flux up to 700 GeV. These results
generated widespread interest and discussions on the origin
of high-energy positrons and electrons~\cite{PsrAMS1}~\cite{crac1}~\cite{DMAMS1}.\\
In this document, based on the published result~\cite{Allele2014}, we 
present a dedicated measurement of the combined (e$^+$ + e$^-$)  up to 1 TeV.\\
The isotropic flux of cosmic rays electrons and positrons in each energy bin $E$, of width $\Delta E$, is given by~\cite{Allele2014}:
\begin{equation}
\Phi(E) = \frac{N_{e}(E)}{A_{eff} \cdot T(E) \cdot \Delta(E)}
\end{equation}\label{flux}
where $N_{e}(E)$ is the number of electrons plus positrons with energy between $E$ and $E+ \Delta E$, $A_{eff}$ is the effective acceptance, $T(E)$ is the exposure time. 
The flux is measured in 74 energy bins from 0.5 to 1 TeV, and the bin width is chosen to be at least two times the energy resolution. The bin-to-bin migration error is 1$\%$ at 1 GeV decreasing to 
0.2$\%$ above 10 GeV. 
The effective acceptance $A_{eff}$ is the product of the detector geometric acceptance ($\sim$~500 cm$^2$sr) and the selection efficiency, estimated with 
simulated events and validated with a pure sample of electron events identified in the data. 
The exposure time is evaluated as a function of energy and it takes into
account the lifetime of the experiment which depends on its orbit
location and on the geomagnetic cutoff~\cite{storm}. 
\begin{figure*}[htb!]
\centering
\includegraphics[width=120mm]{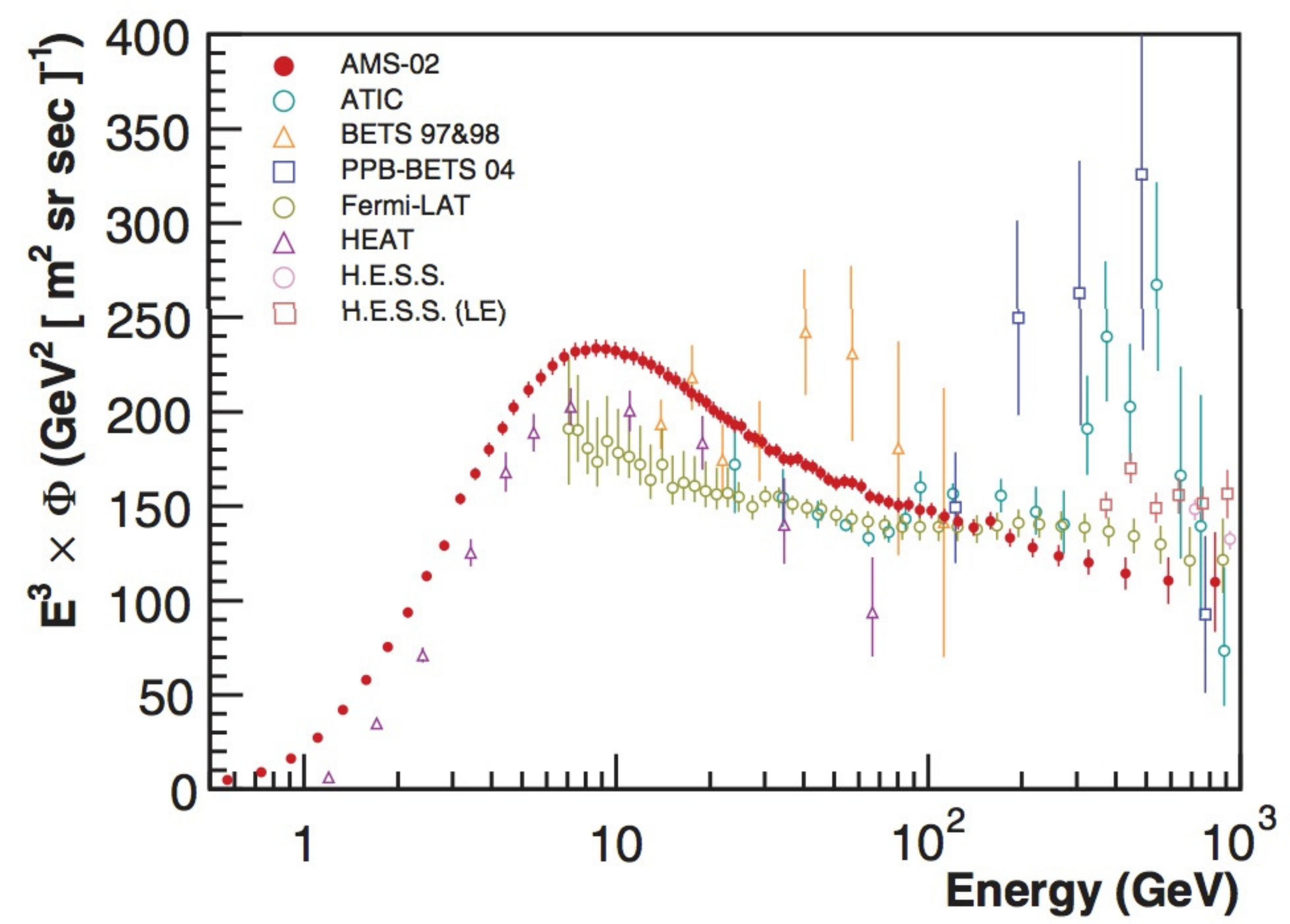}
\caption{AMS-02 combined positrons plus electrons flux, rescaled by the cube of energy, as a function of energy~\cite{Allele2014}, together with the most recent measurements from other experiments. See ~\cite{Allele2014} for the complete list of references}\label{ecrs_2016_alle}
\end{figure*}
To identify downward-going particles of charge one, cuts are applied on the velocity measured by the TOF and 
on the charge reconstructed by the tracker and the TRD. The energy deposited in the ECAL is also used to reject events compatible with a minimum ionising particle.   
To reject positrons and electrons produced by the interaction of primary cosmic rays with the atmosphere, the minimum energy within the bin is required to exceed 1.2 times the geomagnetic cutoff.
Over a sample of well reconstructed particles with a single track in the tracker passing through the ECAL, 
 the identification of signal events is 
performed by first applying  a fixed cut in the ECAL estimator to further reduce the proton background.
The number of signal and background event is estimated for each energy bin performing a template fit procedure, described in ~\cite{Allele2014}.
In total, 10.6 $\times 10^6$ events are identified as electrons and positrons with energies from 0.5 GeV to 1 TeV.
A major experimental advantage of the combined flux analysis compared to the measurement of the individual positron and electron fluxes, especially at high energies, is that the selection does not depend on the sign
of the charge, implying higher selection efficiency. Consequently, this measurement is extended to 1~TeV with less overall uncertainty over the entire energy range.\\ 
The absolute energy scale is verified using minimum ionizing particles and the ratio between the energy, measured by the ECAL, and the momentum, measured by the tracker. 
These results are compared with the test beam values where the energy beam is known to high precision. Between 10 and 290 GeV (Test Beam energies), the 
uncertainty on the absolute scale is $\sim 2\%$, while below 10 GeV it increases to 5 $\%$ at 0.5 GeV and above 290 GeV to 5 $\%$ at 1 TeV. This is treated as an uncertainty on the bin boundaries. 
The statistical error dominates above 140 GeV, as it is shown in the table 1 in~\cite{Allele2014}.\\
Figure~\ref{ecrs_2016_alle} shows the AMS-02 combined electrons plus positrons flux, rescaled by the cube of the energy, as a function of energy, together with previous measurements.
Given the high statistics and high precision of the measurement, the spectral index of the combined flux
has also been measured and, for energies higher than 30 GeV, it is found to be compatible with a single power law.\\
The measurement of the separate fluxes~\cite{epflux2014} shows that electron and positron fluxes 
are different in magnitude and in their energy dependence. 
Above 20 GeV, the positron flux is significantly harder than the electron 
flux, implying that the observed rise in the positron fraction is  due to an excess of positrons and not to a loss of electrons.
This indicates that high energy positrons have a different origin from that of electrons.  
To quantitatively study the energy dependence of the flux in a model independent way, the flux is fit with a spectral index $\gamma$ as: $\Phi(e^+ + e^-)= C E^{\gamma} $
over a sliding energy window. The energy E is expressed in GeV and C is the normalisation constant. 
The resulting energy dependence of the fitted spectral index is shown in figure~\ref{gamma}, where the shading indicates the correlation between the neighbouring points due to sliding energy window. 
Interestingly, the combined $(e^+ + e^-)$ flux  can be described by a single power law above 30.2 GeV, with $\gamma = -3.170 \pm 0.008 (stat+syst) \pm 0.008(energy~scale)$.
The flux, measured from 0.5 to 1 TeV, is smooth and reveals new and distinct information. No structures were observed. 
\begin{figure*}[htb]
\centering
\includegraphics[width=110mm]{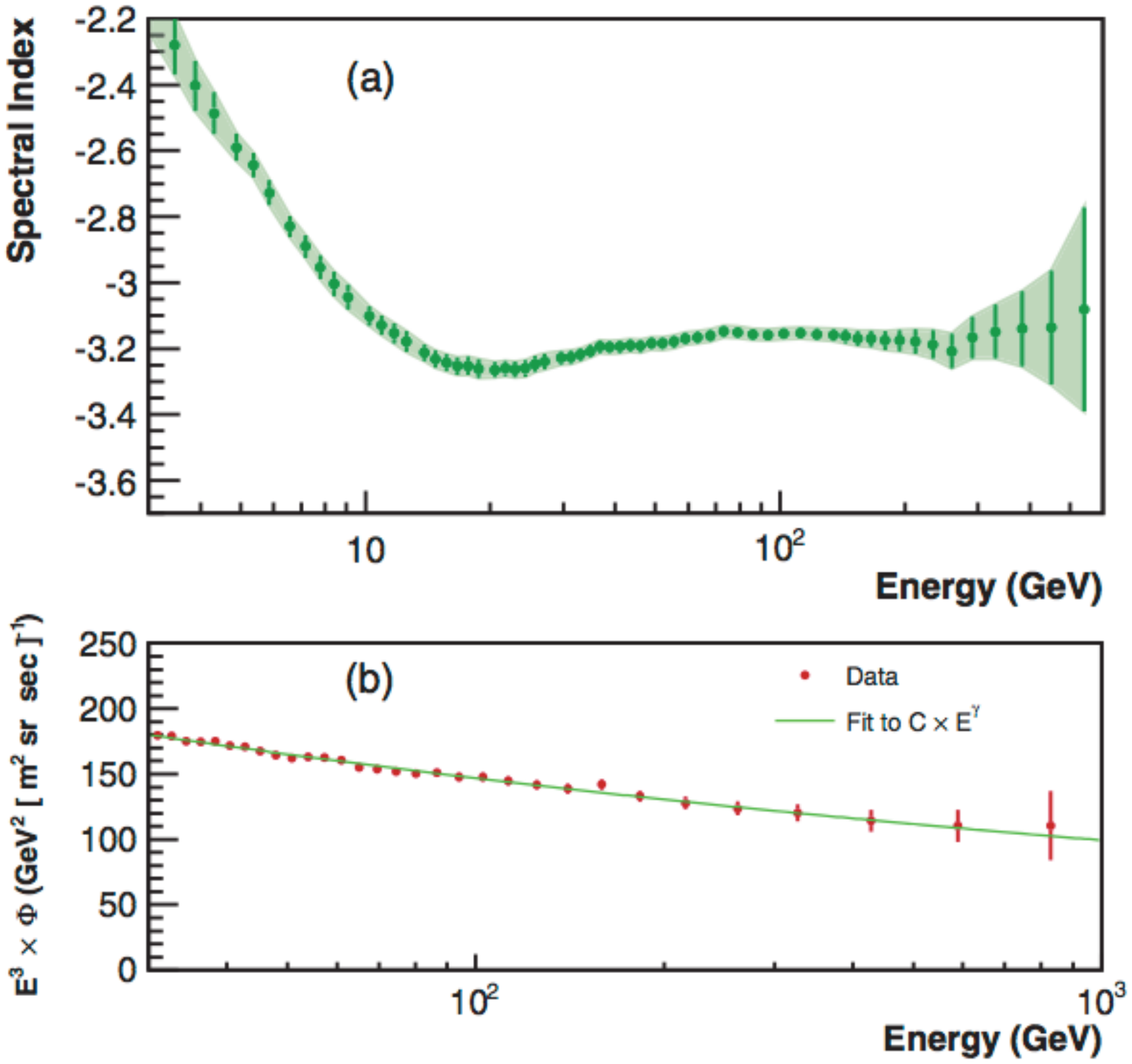}
\caption{(a) the spectral index of the $\Phi(e^+ + e^-)$ flux as a function of energy. The shaded regions indicate the 68$\%$ C.L. intervals including the correlation between neighboring points due to the sliding energy window.
(b) $\Phi(e^+ + e^-)$ multiplied by the cube of the energy as function of energy, above 30.2 GeV, together with the fit to a single power law. 
}\label{gamma}
\end{figure*}

\bigskip 
\begin{acknowledgments}
The author is grateful to the S\~ao Paulo Research Foundation (FAPESP) for the financial support (grant
n. 2014/19149-7 and n. 2014/50747-8). The author is thankful to the organisers of the ECRS conference for their kind availability. 
\end{acknowledgments}

\bigskip 

\end{document}